\documentclass[aps,pra,reprint,groupedaddress,amsmath,amssymb]{revtex4-1}
\usepackage{bm}
\usepackage{graphicx}
\usepackage[usenames, dvipsnames]{color}
\usepackage{subcaption}
\graphicspath{{figures/}}

\DeclareMathOperator{\tr}{tr}
\newcommand*\diff{\mathop{}\!\mathrm{d}}
\newcommand*\Diff{\mathop{}\!\mathrm{D}}
\usepackage{tikz}
\usepackage{ifthen}
\usepackage{tikz-3dplot}
\usepackage{pgfplots}
\usepackage{adjustbox}
\usepgfplotslibrary{colorbrewer}

\begin{document}
\title{Large Deviations
and Fluctuation Theorem for the Quantum Heat Current in the Spin-Boson
Model}%
\author{Erik Aurell}%
\email{eaurell@kth.se}
\affiliation{KTH – Royal Institute of Technology, AlbaNova University Center, SE-106 91 Stockholm, Sweden}%
\affiliation{
Faculty of Physics, Astronomy and Applied Computer Science, Jagiellonian University, 30-348 Krak\'ow, Poland
}
\author{Brecht Donvil}%
\email{brecht.donvil@helsinki.fi}
\affiliation{Department of Mathematics and Statistics, University of Helsinki, P.O. Box 68, 00014 Helsinki, Finland}
\author{Kirone Mallick}%
\email{kirone.mallick@ipht.fr}
\affiliation{Institut de Physique Th\'{e}orique, Universit\'{e} Paris-Saclay, CEA and CNRS, 91191 Gif-sur-Yvette, France}%
\date{October, 2019}
\begin{abstract}
  We study the heat current flowing between two baths consisting of
harmonic oscillators interacting with a qubit through a spin-boson coupling.
An explicit expression for the  generating function of  the total  heat
flowing between the right and left baths is derived by
evaluating the corresponding
Feynman-Vernon path integral by  performing the
non-interacting blip approximation (NIBA). We recover the known expression, obtained by using the polaron transform.
This  generating function  satisfies the
Gallavotti-Cohen fluctuation theorem, both
before and after performing the NIBA. 
We also verify that the heat conductance is proportional to the variance
of the heat current, retrieving the well known fluctuation dissipation relation.
Finally, we present numerical results for the heat current.
\end{abstract}
\maketitle
\section{Introduction}

The flow of a non-vanishing macroscopic current of energy,
charge, matter or information, that breaks time-reversal invariance, 
is a  fingerprint of non-equilibrium behavior. A 
paradigmatic model  for such a  situation consists in a  small system, with
a finite number of degrees of freedom,  that connects   two large reservoirs
in different thermodynamic states. The ensuing stationary state  can not
be described by the standard laws of thermodynamics:
in particular, the steady-state statistics is not given by a Gibbs ensemble.
The theoretical  analysis of simple
models, whether classical or quantum,  provides us
with a wealth of information about far-from-equilibrium physics and has
stimulated numerous studies in the last two decades
\cite{KamenevBook,WeissBook,HaRa2006,open,Rivas,MiPl17,Derrida2007,Gianni2014}.

Quantum systems based on nano-scale integrated circuits are very
effective for the study of quantum phenomena and are  good candidates
for possible applications. This is due to their macroscopic size and
the ensuing ability to manipulate them. For any application minimizing
or controlling the heat flow is essential. Therefore there has been a
great deal of experimental
\cite{PeSo13,PekolaNature2015,GaVi2015,ViAn2016,PeBa2017,KaPek2017,Ronzani2018,SeGu2019}
and theoretical interest
\cite{Gasparinetti1,CampisiHeatEngines,Donvil2018,DonvilCal} in studying the
heat flow in such circuits.  The vast majority of  theoretical
studies  has been focused on the weak coupling regime, for which
well-controlled approximation schemes are available. For
example, in the case of a small system interacting with an environment, it is
possible to integrate out the bath from the full dynamics and express
the resulting system dynamics in terms of a Lindblad equation
\cite{LI76,open,HaRa2006}. In this case, heat currents
can be studied in terms of energy changes of the
system. However, the weak coupling  assumption
deviates from exact treatments
quantitatively and qualitatively  already at 
moderately low couplings \cite{TuSt2019}.

There have been various earlier studies in the strong coupling
regime. 
Based on the polaron transform,  the authors of
\cite{Segal2005} obtained an
analytical expression for the heat current through an $N$ level
system. The polaron transform provides a shortcut for the Non-Interacting Blip Approximation (NIBA) \cite{Aslangul1986,Dekker1987}. The full generating function for a spin-boson system was derived, using the polaron transform, in \cite{Nicolin2011,NiSe2011} and reviewed in \cite{Friedman2018}. The authors of \cite{Wang2015,Wang2017,Liu2018} derived a non-equilibrium polaron-transformed Redfield equation that unifies strong and weak-coupling behavior.  
In \cite{Carrega2016,Motz2018,Aurell2019} the authors start
from  the generating function of the heat current to study its first
moment. Numerical studies include simulations based on  hierarchical
equations of motion 
\cite{Tanimura1989,Tanimura2014,Tanimura2015,Kato2015,Kato2016,deVega2017} 
the quasi-adiabatic propagator path integral (QuAPI)\cite{Makri1998,Boudjada2014},
the iterative full
counting statistics path integral \cite{Kilgour2019},
the multi-configuration time-dependent Hartree (MCTDH) approach~\cite{Velizhanin2008},
the Stochastic Liouvillian algorithm~\cite{StockburgerMak99}, and other Monte Carlo approaches~\cite{Saito2013}.
Other recent
contributions are \cite{Esposito2015,Newman2017,Ronzani2018,Dou2018,
Eisert18,KwUm2018,AurellKawai}.

In this paper we consider a qubit coupled to two (or more) thermal
baths. We derive the full generating function for the spin-boson by directly applying the non-interacting blip approximation (NIBA), without passing through the polaron transform as was done in the original derivation \cite{Nicolin2011,NiSe2011}. We show that we recover the result by \cite{Nicolin2011,NiSe2011}.
Following \cite{Aurell2019}, this generating
function can be written explicitly as 
Feynman-Vernon type path integral. Relying on a
modified version of the NIBA,
an expression for the first moment of the
generating function, i.e. the average heat current, was obtained by directly applying the NIBA in
\cite{Aurell2019}.
Furthermore, we discuss the Gallavotti-Cohen
 relation (see \cite{EvansCM,GCohen1,LeSpo1999,Maes,Derrida2007,Gaspard2009} and references therein), which holds before after the NIBA, as derived in \cite{Nicolin2011,NiSe2011}, in terms of an explicit time reversal.
 Finally, we find a
fluctuation-dissipation relation between the variance of the heat
current and the thermal conductance.

The paper is structured as follows. In section \ref{sec:model}, we
briefly introduce the spin-boson model that we shall analyze. In section
\ref{sec:Generating-function}, the generating function of the
heat current is calculated after performing the NIBA approximation.
 In section
\ref{sec:G-C} we discuss the Gallavotti-Cohen relation before and
after the NIBA. In section \ref{sec:fluc-diss} we invert the
Laplace transform of the generating functions for small $\alpha$ and
obtain a fluctuation-dissipation relation between the variance of the
heat current and the heat conductance. Finally, in section \ref{sec:numerical} we numerically evaluate the first moment of the generating function. Technical details are provided
in the appendices.

\section{The model}\label{sec:model}

The spin-boson model is a prototype for understanding quantum
 coherence in presence of dissipation
  \cite{Chakra82,Bray82,Leggett84,Hakim1985,Leggett1987}.
 It can be viewed as variant of the Caldeira-Leggett model in which
 a quantum particle interacts with a bath of quantum-mechanical oscillators.
 In the spin-boson model,  a two-level system modeled by a spin-1/2
 degree of freedom is put in contact with one or more heat-baths.
 The literature in the subject is vast and we refer the reader to
 some  reviews and to the references therein
 \cite{Leggett1987,Cedraschi2001,LeHur2008,WeissBook}.
 
 In this paper,  we  shall study  two  baths
 made of harmonic oscillators that interact 
 with a qubit via the spin-boson interaction.
 Although there  is no direct interaction between the baths,  energy
  will be  transferred through the qubit.
  The Hamiltonian governing the total evolution of the qubit and
   of the baths is given by 
\begin{equation}
H=H_S+H_L+H_R+H_{LS}+H_{RS}.
\end{equation}
The qubit Hamiltonian is given by
\begin{equation}
H_S=-\hbar\frac{\Delta}{2}\sigma_x+\frac{\epsilon}{2}\sigma_z.
\end{equation}
The left bath and right bath Hamiltonians are given by
\begin{equation}
H_L=\sum_{b\in C}\frac{p_{b,L}^2}{2m_{b,L}}+\frac{1}{2}m_{b,L} \omega_{b,L}^2q_{b,L}^2
\end{equation}
\begin{equation}
H_R=\sum_{b\in R}\frac{p_{b,R}^2}{2m_{b,R}}+\frac{1}{2}m_{b,R} \omega_{b,R}^2q_{b,R }^2.
\end{equation}
Finally, the system-bath interactions are of the spin-boson type
\cite{Leggett1987}
\begin{equation}
H_{LS}=-\sigma_z\sum_{b\in L}C_{b,L}q_{b,L}
\end{equation}
\begin{equation}
H_{RS}=-\sigma_z\sum_{b\in H}C_{b,R}q_{b,R} .
\end{equation}

The effects of the environment are embodied in the spectral
density of the environmental coupling \cite{WeissBook}
(one for each bath):
\begin{equation}
  J^{R/L}(\omega)=\sum_{b\in R/L}\frac{(C^{R/L}_b)^2}{2m_b\omega_b}
  \delta(\omega - \omega_b) \, .
  \label{OhmicJ}
\end{equation}
We shall assume a Ohmic spectrum with an exponential cut-off determined by the frequency $\Omega$
\begin{equation}\label{eq:def-expcutoff}
  J^{R/L}(\omega)=\frac{2}{\pi} \eta_{R/L} {\omega}\exp\left(-\frac{\omega}{\Omega}\right)
\end{equation}
We denote by $U_t$ the unitary evolution operator of the total system 
and assume that the baths are initially at thermal equilibrium
and are prepared in Gibbs states at different temperatures.
For an initial state of the qubit $|i\rangle$ and a final state $|f\rangle$,
the generating function
of the heat current is defined as 
\begin{align}\label{genbath}
  G_{i,f}(\vec{\alpha},t)=&\tr_{R,L}\langle f|e^{i(\alpha_RH_R+\alpha_L H_L)/\hbar}U_te^{-i(\alpha_RH_R+\alpha_L H_L)/\hbar}\nonumber\\&\qquad\times
  \Big(\rho_{\beta_L}\otimes\rho_{\beta_R}\otimes|i\rangle \langle i| \Big)
   \, U^\dagger_t|f\rangle,
\end{align}
with $\vec{\alpha}=(\alpha_R,\alpha_L)$. The trace is taken over all
the degrees of freedom
of the baths. This generating function will allow us to calculate all
the moments of the heat current: for example, by taking the first derivative of $\alpha_L$ and setting $\vec{\alpha}$ to zero gives the change in expected energy of the cold bath
\begin{eqnarray}
-i\hbar\partial_{\alpha_L}G_{i,f}(\vec{\alpha})\big|_{\vec{\alpha}=0}&=&\text{tr}_{R,L}(H_L\rho(t))-\text{tr}(H_L\rho(0)) \nonumber \\ &=&\Delta E_L.
\end{eqnarray}

\section{Calculation of the generating function}\label{sec:Generating-function}

The full generating function \eqref{genbath} was calculated in \cite{Nicolin2011,NiSe2011} using the polaron transform. In this section we aim to perform this calculation by explicitly applying the NIBA to \eqref{genbath}.
The first step of the calculation
is to rewrite the  trace in equation \eqref{genbath}  as a Feynman-Vernon type path-integral \cite{WeissBook}. After integrating over the left
and the right bath, see Appendix \ref{sec:appGenn}, the expression for the generating function
is given by \cite{Aurell2019}
\begin{align}\label{genNobath}
&G_{i,f}(\vec{\alpha},t)=\int_{i,f}\Diff X\Diff Ye^{\frac{i}{\hbar}S_0[X]-\frac{i}{\hbar}S_0[Y]}\mathcal{F}_{\vec{\alpha}}[X,Y],
\end{align}
where $\mathcal{F}_{\vec{\alpha}}$ is the influence functional.
The paths $X$ and $Y$ are the forward and backwards path of the qubit,
they take values $\pm1$. The forward path $X$ corresponds to the forward evolution operator $U_t$ in \eqref{genNobath}, and $Y$ corresponds to $U^\dagger_t$. In the absence of interactions with the baths, the dynamics of the qubit are fully described by the free qubit action $S_0$
\begin{equation}
S_0[X]=-\frac{\epsilon}{2}\int dt\, X(t)-i\log(i\Delta dt /2)\int |dX(t)|
\end{equation}
The integral $\int |dX(t)|$ counts the amount of jumps in the path. Thus, when the path $X$ makes $n$ jumps, the second term gives the weight $(i\Delta dt /2)^n$.
 The
effect of the influence functional is to  generate
interactions between the forward and backward paths; it also
embodies  the dependence on the parameters  $\vec{\alpha}$.
\begin{eqnarray}\label{eq:influence}
  \mathcal{F}_{\vec{\alpha}}[X,Y]=e^{\frac{i}{\hbar}(S_{i,\alpha_L}^C[X,Y]+S_{i,\alpha_R}^R[X,Y])}
  \nonumber \\
  \qquad\times  e^{ -\frac{1}{\hbar}(S_{r,\alpha_L}^C[X,Y]+S_{r,\alpha_R}^R[X,Y])},
\end{eqnarray}
where  the real part of the interaction action is given by \cite{AurKaw2019}
\begin{align}\label{eq:Sr}
&S_{r,\alpha_{R/L}}^{R/L}[X,Y]=\int_{t_i}^{t_f}\diff t\int_{t_i}^{t}\diff s \bigg((X_tX_s+Y_tY_s)k^{R/L}_r(t-s)\nonumber\\&-X_tY_sk^{R/L}_r(t-s+\alpha_{R/L})-X_sY_tk^{R/L}_r(t-s-\alpha_{R/L})\bigg)
\end{align}
and the imaginary part is defined as 
\begin{align}\label{eq:Si}
  & S_{i,\alpha_{R/L}}^{R/L}[X,Y]=\int_{t_i}^{t_f}\diff t\int_{t_i}^{t}\diff s \bigg((X_tX_s-Y_tY_s)k^{R/L}_i(t-s)\nonumber\\
  &+X_tY_sk^{R/L}_i(t-s+\alpha_{R/L})-X_sY_tk^{R/L}_i(t-s-\alpha_{R/L})\bigg)
\end{align}
The kernels that appear in these expressions are 
\begin{equation}
k_i^{j}(t-s)=\sum_{b}\frac{(C^{j}_{b,j})^2}{2m_{b,j}\omega_{b,j}}\sin(\omega_{b,j}(t-s))
\end{equation}
and
\begin{align}
k_r^{j}(t-s)=\sum_{b}&\frac{(C^{j}_{b,j})^2}{2m_{b,j}\omega_{b,j}}\coth\left(\frac{\hbar\omega_{b,j}\beta_{j}}{2}\right)\nonumber\\&\times\sin(\omega_{b,j}(t-s)),
\end{align}
for $j=R,L$.
The integral of the bath degrees of freedom being performed, the
generating function is given as the 
qubit path-integral  \eqref{genNobath} over two binary paths. This
remaining expression can not be calculated exactly; in the next section,
we shall evaluate the generating function
by resorting to  the Non-Interacting Blip Approximation
(NIBA).

\subsection{Performing the NIBA}\label{subsec:niba}

Originally,  the  idea of the NIBA was proposed  in \cite{Chakravarty1984}, see also \cite{Leggett1987},
to compute transition probabilities between states of the qubit: this
corresponds to taking $\alpha=0$ in (\ref{genbath}).
The paths $X$ and $Y$ being binary, there are  only two possibilities
at a given time: either  $X = Y$, this is a {\it Sojourn}  or
 $X = - Y$, this is a {\it Blip}. The NIBA
 approximation 
 relies on two assumptions (explained in  \cite{Leggett1987}):

 (i) The typical Blip-interval time $\Delta 
 t_B$ is much shorter than the typical Sojourn-interval time $\Delta t_S$: $\Delta t_B\ll\Delta t_S$.

 (ii) Bath correlations decay over times much smaller than the typical Sojourn interval $\Delta t_S$.
 
For an Ohmic spectrum \eqref{eq:def-expcutoff}, these assumptions are valid for two regimes: (a) for $\epsilon=0$ and weak coupling and (b) for large damping and/or at high temperatures \cite{WeissBook}.

 Under these assumptions, the only
 nonzero contributions to the time integrals in the interaction part of the action \eqref{eq:Sr} and \eqref{eq:Si} are obtained when
\begin{itemize}
\item[a.] $t$ and $s$ are in the same Blip-interval
\item[b.] $t$ and $s$ are in the same Sojourn-interval
\item[c.] $t$ is in a Sojourn and  $s$ is an adjacent Blip interval
 \item[d.] $t$ and $s$ are both in Sojourn-intervals separated by one Blip.
\end{itemize}
Other terms cannot contribute since then $t$ and $s$ will be situated in intervals separated by at least one Sojourn, which does give a contribution under assumption (ii).

The strategy to perform the NIBA is to break up the integrals \eqref{eq:Sr} and
\eqref{eq:Si} over the whole time interval into a sum of the surviving
parts, which can be evaluated separately.

\begin{figure}
\includegraphics[scale=1]{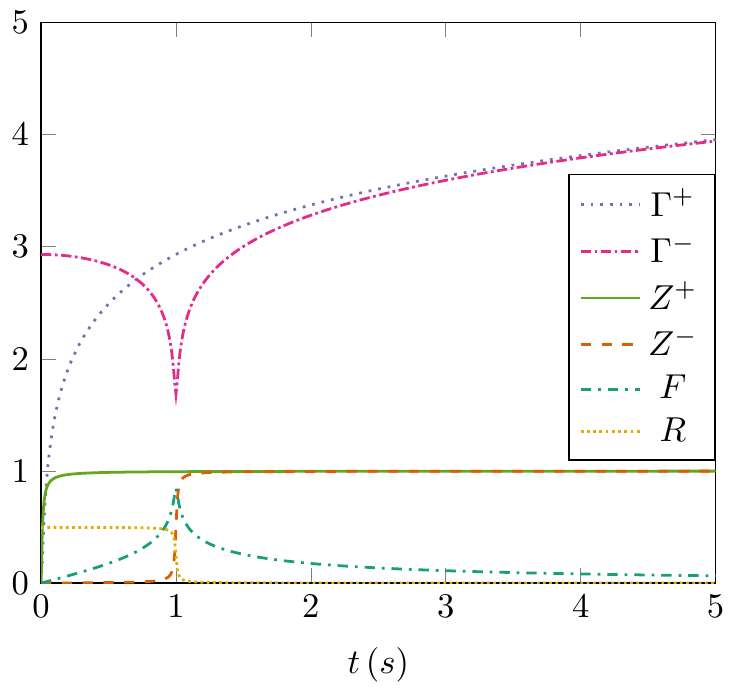}
\caption{Behaviour of the functions \eqref{eq:def-functions} appearing in the definitions of the matrix elements \eqref{eq:def-matrixelements} for $\alpha=1$, $\eta=1$ and $\beta=(0.1\,\text{K}k_B)^{-1}$.}
\label{fig:functions}
\end{figure}

In the present work, we
extend the NIBA to include nonzero $\alpha$ (see also \cite{Aurell2019}),
which leads to a time shift
in some of the Kernels in the action \eqref{eq:Sr} and
\eqref{eq:Si}. In the framework of our  approximation, we consider values of
$\alpha_{R/L}$, such that  $\alpha_{R/L}\ll \Delta t_S$. Following the
same reasoning as for $\alpha_{R/L}=0$, it is clear that under said
additional assumption, the same terms as before have a chance of being
nonzero. In Appendix \ref{app:NIBA},  we explicitly calculate the five different
surviving terms \footnote{For c. the Blip can be before or after the Sojourn. These give different contributions and are calculated separately. Hence there are five different surviving terms.} after the NIBA. The resulting expression for the
generating function can be written in terms of a transfer matrix
$M(\alpha,t)$:
\begin{align}\label{generatinTransf}
&G_{\uparrow \uparrow}(\vec\alpha,t)+G_{\uparrow\downarrow}(\vec\alpha,t)=\begin{pmatrix}
1&1
\end{pmatrix}\sum_{n=0}^{+\infty}(-1)^n\left(\frac{\Delta}{2}\right)^{2n}\nonumber\\&\times\int \diff t_1\hdots \diff t_{2n}\mathbf{M}(\alpha,\Delta t_{2n})\mathbf{M}(\alpha,\Delta t_{2n-2})\hdots\mathbf{M}(\alpha,\Delta t_2)\begin{pmatrix}
1\\0
\end{pmatrix}
\end{align}
where $\Delta t_{2j} = t_{2j} - t_{2j-1}$. The  
 transfer matrix $\mathbf{M}$  is given by 
\begin{align}\label{eq:transf}
&\mathbf{M}(\vec\alpha,t)=  2\begin{pmatrix}
A(t) &-B(\vec\alpha,t)\\-C(\vec\alpha,t)&D(t)
\end{pmatrix}
\end{align}
Note that only the off-diagonal elements of the transfer matrix depend on $\alpha$.
The functions $A,B,C$ and $D$ that appear as matrix elements in $\mathbf{M}$
are determined once the NIBA has been performed. Their values are given by
\begin{subequations}\label{eq:def-matrixelements}
\begin{equation}
A(t)=\cos\frac{1}{\hbar}\left(Z_L^+(t)+Z_R^+(t)-\epsilon t\right)e^{-\frac{1}{\hbar}(\Gamma_L^+(t)+\Gamma_R^+(t))} 
\end{equation}
\begin{align}
&B(\vec\alpha,t)=e^{-\frac{1}{\hbar}(\Gamma_L^-(\alpha_L,t)+\Gamma_R^-(\alpha_R,t)+2i(R_L(\alpha_L,t)+R_R(\alpha_R,t))}\nonumber\\&\times \cos\frac{1}{\hbar}\big(Z_L^-(\alpha_L,t)+2iF_L(\alpha_L,t)+Z_R^-(\alpha_L,t)\nonumber\\&\qquad+ 2i F_R(\alpha_L,t)+\epsilon t\big)
\end{align}
\begin{align}&C(\vec\alpha,t)=e^{-\frac{1}{\hbar}(\Gamma_L^-(\alpha_L,t)+\Gamma_R^-(\alpha_R,t) +2 i(R_L(\alpha_L,t)+R_R(\alpha_R,t))}\nonumber\\&\times \cos\frac{1}{\hbar}\big(Z_L^-(\alpha_L,t)+2iF_L(\alpha_L,t)+Z_R^-(\alpha_L,t)\nonumber\\&\qquad+ 2i F_R(\alpha_L,t)-\epsilon t\big)
\end{align}
\begin{equation}
D(t)=\cos\frac{1}{\hbar}\left(Z_L^+(t)+Z_R^+(t)+\epsilon t\right)e^{-\frac{1}{\hbar}(\Gamma_L^+(t)+\Gamma_R^+(t))} 
\end{equation}
\end{subequations}
All  the auxiliary functions
$Z_j^{\pm},\Gamma_j^{\pm}, R_j$ and $F_j$, where the index $j=L,R$
refers  to the left or the right bath, are  determined  in the 
Appendix \ref{app:NIBA}.  Assuming a Ohmic spectral density
with exponential cut-off with frequency $\Omega$ (\ref{OhmicJ}),
 the explicit  expressions of these functions are  given  in the following
equations:
\begin{subequations}\label{eq:def-functions2}
\begin{align}
  &Z^+_{j}(t)=\frac{2\eta_j}{\pi}\int_0^\infty\diff \omega\,
  \frac{\sin(\omega t)}{\omega}e^{-\omega/\Omega}\\
  &Z_j^-(\alpha_j,t)=\frac{2\eta_j}{\pi}\int_0^\infty\diff \omega\,
  \frac{\sin(\omega t)}{\omega}\cos(\omega\alpha_j)e^{-\omega/\Omega}
\end{align}
\end{subequations}
\begin{subequations}\label{eq:def-functions}
\begin{align}
  &\Gamma^+_j(t)=\frac{2\eta_j}{\pi}\int_0^\infty\diff \omega\,
  \frac{1-\cos(\omega t)}{\omega}\coth\left(\frac{\omega\hbar\beta_j}{2}\right)e^{-\omega/\Omega}\\
 & \Gamma^-_j(\alpha_j,t)=\frac{2\eta_j}{\pi}\int_0^\infty\diff \omega\,
 \bigg( \frac{1-\cos(\omega t)\cos(\omega\alpha_j)}{ \omega}\nonumber\\
 &\hspace{2.5cm}\times \coth(\frac{\omega\hbar\beta_j}{2})\,e^{-\omega/\Omega}\bigg)
\end{align}
\end{subequations}
\begin{subequations}
\begin{align}
  &R_j(\alpha_j,t)=\frac{\eta_j}{\pi}\int_0^\infty\diff \omega\,
  \frac{\sin(\omega\alpha_j)}{ \omega}\cos(\omega t)e^{-\omega/\Omega}\\
 & F_j(\alpha_j,t)= \frac{\eta_j}{\pi}\int_0^\infty\diff \omega\,
   \frac{\coth\left(\frac{\omega\hbar\beta_j}{2}\right)}{ \omega}
  \sin(\omega t)\sin(\omega\alpha_j)e^{-\omega/\Omega}
\end{align}
\end{subequations}
The behaviour of these functions is shown in figure \ref{fig:functions}.

 We shall denote by $\tilde{\phi}$ the Laplace transform of a function
 $\phi(\vec\alpha, t)$, defined  as follows: 
 \begin{equation}
\tilde{\phi}(\vec\alpha,\lambda)=\int_0^{\infty}dt\, e^{-\lambda t} \phi(\vec\alpha,t).
 \end{equation}
 Then,  taking the Laplace transform  of \eqref{generatinTransf} leads us to
\begin{align}
&\tilde{G}_{\uparrow \uparrow}(\vec\alpha,\lambda)+\tilde{G}_{\uparrow\downarrow}(\vec\alpha,\lambda)=\nonumber\\&\quad\lambda^{-1}\begin{pmatrix}
1&1
\end{pmatrix}\left(\sum_{n=0}^{+\infty}(-1)^n\left(\frac{\Delta}{2}\right)^{2n}\lambda^{-n}\mathbf{\tilde{M}}^n(\alpha,\lambda)\right)\begin{pmatrix}
1\\0
\end{pmatrix}
\end{align}
We call  $\lambda_+(\vec\alpha,\lambda)$ and $\lambda_-(\vec\alpha,\lambda)$  the eigenvalues of the 2 by 2 matrix 
$\mathbf{\tilde{M}}(\vec\alpha,\lambda)$, with corresponding left eigenvectors $v_+(\vec\alpha,\lambda)$ and $v_-(\vec\alpha,\lambda)$ and right eigenvectors $w_+(\vec\alpha,\lambda)$ and $w_-(\vec\alpha,\lambda)$. We can write $\mathbf{\tilde{M}}^n$ in terms of it eigenvectors and eigenvalues as
\begin{equation}
\mathbf{\tilde{M}}^n=\lambda^n_+(\vec{\alpha},t) w_+(\vec{\alpha},t) v^T_+(\vec{\alpha},t) +\lambda^n_-(\vec{\alpha},t) w_-(\vec{\alpha},t) v^T_-(\vec{\alpha},t) 
\end{equation}
We have 
\begin{align}
\lambda_\pm(\vec\alpha,\lambda) &=
\tilde A(\lambda) + \tilde D(\lambda)  \\ 
& \pm \sqrt{(\tilde A(\lambda)-\tilde D(\lambda)^2+4 \tilde B(\vec\alpha,\lambda)\tilde C(\vec\alpha,\lambda)}\nonumber
\end{align}
Finally, the 
Laplace transform of the generating function takes a simpler form in the
 eigenbasis of $\mathbf{\tilde{M}}$:
\begin{align}\label{eq:lapeig}
&\tilde{G}_{\uparrow \uparrow}(\vec\alpha,\lambda)+\tilde{G}_{\uparrow \downarrow}(\vec\alpha,\lambda)\nonumber\\&\qquad=\frac{Q_+(\vec\alpha,\lambda)}{\lambda+\left(\frac{\Delta}{2}\right)^2\lambda_+(\vec\alpha,\lambda)}+\frac{Q_-(\vec\alpha,\lambda)}{\lambda+\left(\frac{\Delta}{2}\right)^2\lambda_-(\vec\alpha,\lambda)},
\end{align}
where we defined the amplitudes
\begin{equation}
Q_\pm=\begin{pmatrix}
1&1
\end{pmatrix}w_\pm v^T_\pm \begin{pmatrix}
1\\0
\end{pmatrix}.
\end{equation}
Using the relations
\begin{subequations}
\begin{align}
2R_j(\alpha_j,t)\pm Z^-(\alpha_j,t)&=Z^+(\alpha_j \pm t)\\
2F_j(\alpha_j,t)\pm \Gamma^-(\alpha_j,t)&=\Gamma^+(\alpha_j \mp t).
\end{align}
\end{subequations}
one can check that equation \eqref{eq:lapeig} recovers the result derived by \cite{Nicolin2011,NiSe2011}

\section{The fluctuation theorem  pre- and post-NIBA}\label{sec:G-C}
The authors of \cite{Nicolin2011} proved a fluctuation relation for the generating function after the NIBA was performed. They showed that the leading eigenvalue of the transfer matrix $\tilde{\mathbf{M}}$ is invariant under $\vec{\alpha}\rightarrow i (\beta_L,\beta_R)-\vec{\alpha}$. Here we show said fluctuation relation directly on $\mathbf{M}$ by considering a proper time reversal.
\subsection{Time Reversal}
There are multiple ways to define a time reversal for a stochastic process, see \cite{Chetrite2008}. In this work, we define a reversal for the qubit state paths $X(t)$ and $Y(t)$ as
\begin{subequations}\label{eq:timerev}
\begin{align}
X_R(t)&=Y(t_f+t_i-t)\\
Y_R(t)&=X(t_f+t_i-t),
\end{align}
\end{subequations}
see figure \ref{fig:rev}. In the time reversed path the forward and backward path interchange and run from $t_f$ to $t_i$.
To illustrate this time reversal, let us note that for $\vec{\alpha}=0$, the generating function \eqref{genbath} can be written as 
\begin{align}
&G_{i,f}(0)=\tr_{R,C}(\rho_{\beta_L}\otimes\rho_{\beta_R}\langle i|U|f\rangle \langle f|U^\dagger|i\rangle)
\end{align}
Expressing the trace as a path integral and computing the trace over the bath variables gives
\begin{align}
&G_{i,f}(\vec{0})=\int_{i,f}\Diff X\Diff Ye^{-\frac{i}{\hbar}S_0[X]+\frac{i}{\hbar}S_0[Y]}\mathcal{F}_R[X,Y],
\end{align}
With influence functional
\begin{equation}
\mathcal{F}_{R}[X,Y]=e^{\frac{i}{\hbar}(S_{i,R}^L+S_{i,R}^R)[X,Y]-\frac{1}{\hbar}(S_{r,R}^L+S_{r,R}^R)[X,Y]},
\end{equation}
where we defined the real part of the action as
\begin{align}
&S_{r,R}^{R/L}[X,Y]=\int_{t_i}^{t_f}\diff t\int_{t_i}^{t}\diff s \bigg((X_tX_s+Y_tY_s)k^{R/L}_r(t-s)\nonumber\\&\quad-X_tY_sk^{R/L}_r(t-s)-X_sY_tk^{R/L}_r(t-s)\bigg)
\end{align}
and the imaginary part
\begin{align}
&S_{i,R}^{R/L}[X,Y]=\int_{t_i}^{t_f}\diff t\int_{t_i}^{t}\diff s \bigg((Y_tY_s-X_tX_s)k^{R/L}_i(t-s)\nonumber\\&\quad+X_tY_sk^{R/L}_i(t-s)-X_sY_tk^{R/L}_i(t-s)\bigg)
\end{align}
Now taking $X(t)\rightarrow X_R(t)$ and $Y(t)\rightarrow Y_R(t)$, retrieves the expression for the generating function \eqref{genNobath} for $\vec{\alpha}=0$.

\subsection{The Gallavotti-Cohen symmetry}
Let us define the time reversed generating function as
\begin{align}\label{eq:genrev1}
&G^R_{fi}(\alpha_R,\alpha_L,t)=\tr_{R,L} \langle i|e^{i(\alpha_RH_R+\alpha_L H_L)/\hbar}U^\dagger_t \nonumber\\&\quad\times e^{-i(\alpha_RH_R+\alpha_L H_L)/\hbar}(\rho_{\beta_L}\otimes\rho_{\beta_R}\otimes|f\rangle \langle f|)U_t|i\rangle.
\end{align}

Before performing the NIBA, it is straightforward to show from the definition of the generating function \eqref{genNobath} and  \eqref{eq:genrev1} that the Gallavotti-Cohen relation holds:
\begin{align}\label{eq:g-c}
&G_{if}(i\beta_R\hbar-\alpha_R,i\beta_L\hbar-\alpha_L,t)= G^R_{fi}(\alpha_R,\alpha_L,t),
\end{align}
see \cite{Andrieux2009} for a detailed discussion on the Gallavotti-Cohen relation for interacting systems.
After integrating out the bath, the above equation can be checked using the time reversal defined in \eqref{eq:timerev}.

\begin{figure}
\centering
\includegraphics[scale=1]{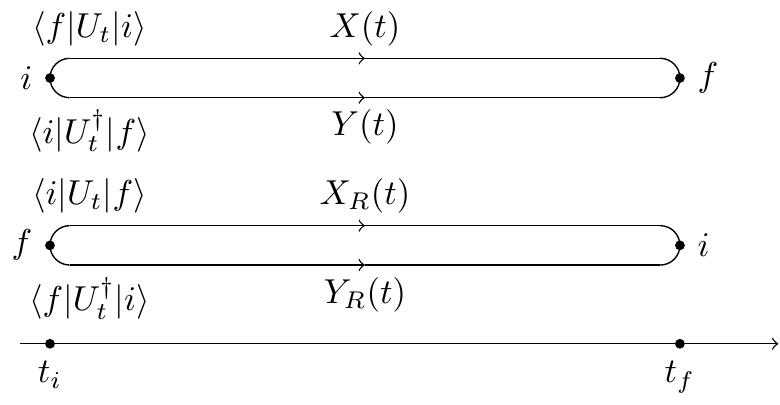}

\caption{Illustration of the (top) forward spin-state paths and the (bottom) time reversed paths.}
\label{fig:rev}
\end{figure}

It is possible to show that the Gallavotti-Cohen relation \eqref{eq:g-c} still holds after performing the NIBA. In order to do so we perform the NIBA on the time-reversed generating function $G^R_{fi}(\alpha_R,\alpha_L,t)$ following the same procedure as outlined in Subsection \ref{subsec:niba}. The result is of the form \eqref{generatinTransf}, with transfer matrix
\begin{align}\label{eq:transfrev}
\bar{\mathbf{M}}(\vec\alpha,t)= 2 \begin{pmatrix}
D(t) &-C(\vec \alpha,t) \\-B(\vec \alpha,t) & A(t)
\end{pmatrix},
\end{align}
on the other hand, one can calculate that
\begin{align}\label{eq:transfiv}
{\mathbf{M}}(i\hbar(\beta_R,\beta_L)-\vec\alpha,t)= 2 \begin{pmatrix}
A(t) &-C(\vec \alpha,t) \\-B(\vec \alpha,t) & D(t)
\end{pmatrix}.
\end{align}
Note that in the time reversal \eqref{eq:timerev}, we interchange the meaning of $X$ and $Y$, as illustrated in figure \ref{fig:rev}. Interchanging the roles of $X$ and $Y$ means flipping the diagonal elements in transfer matrix. Thus ${\mathbf{M}}(i\hbar(\beta_R,\beta_L)-\vec\alpha,t)$ and $\bar{\mathbf{M}}(\vec\alpha,t)$  are equivalent, proving that the Gallavotti-Cohen relation remains true after the performing the NIBA, as was shown by \cite{Nicolin2011}.

\section{Fluctuation-dissipation relation}\label{sec:fluc-diss}
In this section we aim to calculate first and second moment of the heat current in the steady state. In steady state we only need to focus on one bath as the magnitude of the heat current is the same for both baths. Therefore, let us set $\alpha_R=0$ and write $\alpha=\alpha_L$. We invert the Laplace transform of the generating function up to second order in $\alpha$. This allows us direct access to the first and second moment of the heat current. In order to be self-contained, in Appendix \ref{sec:thermal-power} we present a derivation of the thermal conductance $\kappa$ \eqref{eq:kappafinal}, which will appear in the fluctuation-dissipation relation.

Concretely, we look for poles of equation \eqref{eq:lapeig}, by constructing a function
 \begin{equation}\label{eq:lamnu}
\lambda(\alpha)=\lambda_0+\lambda_1\alpha+\lambda_2\alpha^2+O(\alpha^3).
\end{equation}
which solves 
\begin{equation}\label{eq:laml}
\lambda(\alpha)+\left(\frac{\Delta}{2}\right)^2\lambda_-(\alpha,\lambda(\alpha))=0
\end{equation}
at all orders in $\alpha$.
Hence for small $\alpha$, we have, in the long time limit
\begin{align}
G_{i,f}(\alpha,t)= &\text{Res}\left(\frac{e^{\lambda t}Q_-(\alpha,\lambda)}{\lambda(\alpha)+\left(\frac{\Delta}{2}\right)^2\lambda_-(\alpha,\lambda)},\lambda(\alpha)\right)\nonumber\\
=&\frac{e^{\lambda(\alpha)t}Q_-(\alpha,\lambda(\alpha))}{1+\left(\frac{\Delta}{2}\right)^2\dot\lambda_-(\alpha,\lambda(\alpha))}
\end{align}
(Note that in the large  time limit  the contribution of the $\lambda_+$ is
      exponentially subdominant).
Keeping in mind that $\lambda_-(0,\lambda)=0$, the zeroth order of equation \eqref{eq:laml} gives.
\begin{equation}
\lambda_0=0.
\end{equation}
Equation \eqref{eq:laml} to the first order in $\alpha$ translates to
\begin{equation}
\lambda_1+\left(\frac{\Delta}{2}\right)^2\lambda'_-(0,\lambda_0)+\left(\frac{\Delta}{2}\right)^2\dot\lambda_-(0,\lambda_0)\lambda_1
\end{equation}
where the accent denotes the derivative to $\alpha$ the first variable and a dot a derivative to $\lambda$. The steady state heat current is given by $-i\hbar\lambda_1$. After some algebra we find that
\begin{equation}
\lambda_1=i\left(\frac{\Delta}{2}\right)^2\frac{p_+\,\pi_\downarrow+p_-\pi_\uparrow}{\hbar(p_++p_-)},
\end{equation}
where we defined
\begin{subequations}
\begin{align}
C_L(t) &= e^{-\frac{1}{\hbar}\Gamma_L^+(t)+ \frac{i}{\hbar}Z^+_L(t)} \\
C_R(t) &= e^{-\frac{1}{\hbar}\Gamma_R^+(t)+ \frac{i}{\hbar}Z^+_R(t)}
\end{align}
\end{subequations}
and $\hat{C}_L(\omega)$, $\hat{C}_R(\omega)$ their Fourier transforms. 
The fractions $p_{\pm}/(p_++p_-)$ give the steady state population for the qubit in the up/down state, with
\begin{subequations}\label{eq}
\begin{align}
p_+&= \int_{-\infty}^\infty dt\,C_L(t)C_R(t)e^{i\epsilon t}\\
p_-&= \int_{-\infty}^\infty dt\,C_L(t)C_R(t)e^{-i\epsilon t},
\end{align}
\end{subequations}
and the power emitted from the up $\pi_{\downarrow}$ and down state $\pi_{\uparrow}$
\begin{subequations}
\begin{align}
\pi_{\uparrow}&=\frac{\hbar}{2\pi}\int_{-\infty}^\infty d\omega\,\omega\hat{C}_L(\omega)\hat{C}_R(\epsilon-\omega)\\
\pi_{\downarrow}&=\frac{\hbar}{2\pi}\int_{-\infty}^\infty d\omega\,\omega\hat{C}_L(\omega)\hat{C}_R(-\epsilon-\omega).
\end{align}
\end{subequations}
The convolution in the first line can be interpreted as the sum over qubit relaxation rates with energy $\omega$ going to the left bath and $-\omega+\epsilon$ to the right bath, and the second line similarly in terms of a qubit excitation \cite{Nicolin2011}. Additionally, we define
\begin{subequations}
\begin{align}
\Sigma^+&=\frac{\hbar^2}{2\pi}\int_{-\infty}^\infty d\omega\,\omega^2\hat{C}_L(\omega)\hat{C}_R(\epsilon-\omega)\\
\Sigma^-&=\frac{\hbar^2}{2\pi}\int_{-\infty}^\infty d\omega\,\omega^2\hat{C}_L(\omega)\hat{C}_R(-\epsilon-\omega)
\end{align}
\end{subequations}
Similarly, an expression can be obtained for $\lambda_2$. In equilibrium, when $\beta_R=\beta_C$, 
\begin{equation}
\lambda_2=-\frac{\Delta^2}{4\hbar^2}\frac{p_-\Sigma^++p_+\Sigma^-+4 \pi_\uparrow \pi_\downarrow}{p_++p_-}.
\end{equation}

Writing the explicit expression for $Q_-(\alpha,\lambda(\alpha))$, straightforward algebra shows that
\begin{equation}
Q_-(\alpha,\lambda(\alpha))=1+O(\alpha^3), 
\end{equation}
and we obtain  that the generating function is given by
\begin{align}
  &G_{i,f}(\alpha)= e^{(\lambda_1\alpha+\lambda_2\alpha^2+O(\alpha^3))t} \times 
 \hskip 3cm \nonumber\\&\bigg\{1  -\left(\frac{\Delta}{2}\right)^2(\dot\lambda_-'(0,0)+\lambda_-''(0,0)\lambda_1)\alpha
   \nonumber\\&+\bigg(\left(\frac{\Delta}{2}\right)^2\dot\lambda_-'(0,0)\bigg)^2\alpha^2-\left(\frac{\Delta}{2}\right)^2\dot\lambda_-''(0,0)\alpha^2+O(\alpha^3)\bigg\}
 \nonumber \end{align}
 The first moment of the heat current is
 \begin{equation}
\langle \Delta E\rangle=-i\hbar t\lambda_1
 \end{equation}
 which correctly leads to the heat current defined in \eqref{eq:thermal-power-AM}.
The variance of the heat current is then given by
\begin{equation}
\text{Var}[\Delta E]=-\hbar^2 t(2\lambda_2-2(\dot\lambda_-'(0,0)+\lambda_-''(0,0)\lambda_1))\lambda_1)+O(t) .
\end{equation}
In equilibrium, $\lambda_1=0$, we find that
\begin{align}
&\lim_{t\rightarrow\infty}\frac{1}{t}\text{Var}[\Delta E]= -2\hbar^2\lambda_2\\
&=\frac{\Delta^2}{2}\frac{p_-\Sigma^++p_+\Sigma^-+4 \pi_\uparrow \pi_\downarrow}{p_++p_-}.
\end{align}
Comparing to \eqref{eq:kappafinal}, we find the following identity
\begin{equation}
\lim_{t\rightarrow\infty}\frac{1}{t}\text{Var}[\Delta E]=2\kappa,
\end{equation}
which proves the  fluctuation-dissipation relation.

\section{Numerical evaluation of the Generating function}\label{sec:numerical}
In this section we numerically study the heat current predicted by \eqref{generatinTransf}, earlier numerical studies on the spin-boson model include e.g. \cite{Segal2005,Segal2006,Segal2008,Nicolin2011,NiSe2011,Friedman2018,Wang2015,Wang2017,Liu2018}.

The heat current \eqref{eq:thermal-power-AM} is completely determined by the functions $Z^+_{L/R}(t)$ and $\Gamma^+_{L/R}(t)$, defined in \eqref{eq:def-functions} and \eqref{eq:def-functions2}.
For the Ohmic spectral density $J(\omega)$ with exponential cut-off \eqref{eq:def-expcutoff}, these functions have analytic solutions \cite{Leggett1987}
\begin{equation}
Z^+_j(t)=\eta_{j} \tan^{-1}(\Omega t)
\end{equation}
\begin{equation}
\Gamma_j^+(t)=\frac{1}{2}\eta_{j}\log\left(1+\Omega^2t^2\right)
+\eta_{j}\log\left(\frac{\hbar\beta_{j}}{\pi t}\sinh\frac{\pi t}{\hbar\beta_j}\right),
\end{equation}
with $j=L,\, R$.

\begin{figure}
\includegraphics[scale=1]{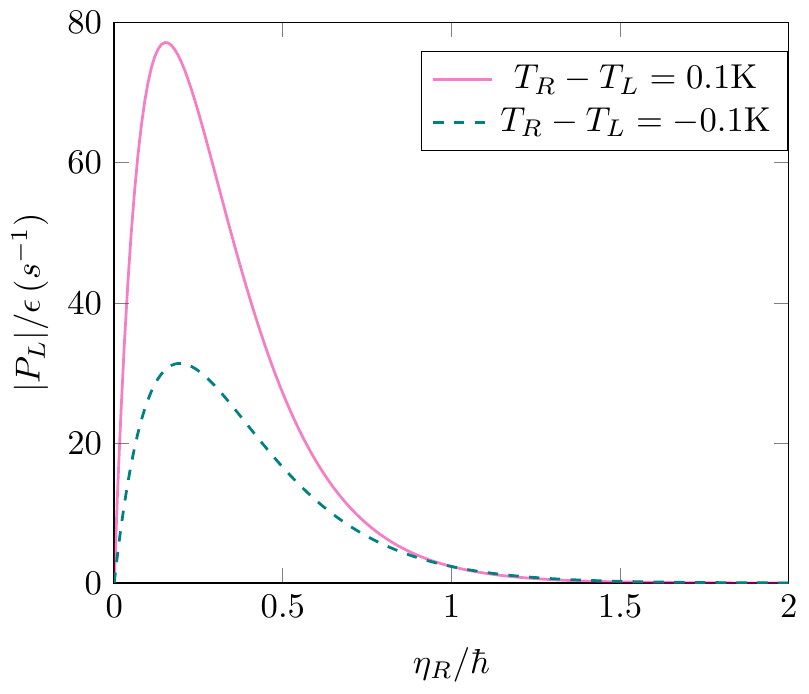}
\caption{Numerical evaluation of the heat current \eqref{eq:thermal-power}. The purple lines are for $T_R=0.2\,\text{K}$, $T_L=0.1\,\text{K}$ and the blue lines for $T_R=0.1\,\text{K}$, $T_L=0.2\,\text{K}$. The other parameters in the model are $\epsilon=1\,\text{K}\times k_B$, $\hbar\Delta=0.01\epsilon$, $\Omega=100\epsilon/\hbar$ and $\eta_L=\hbar$.}
\label{fig:power}
\end{figure}

For our numerical analysis we consider the parameters $\epsilon=1\,\text{K}\times k_B$, $\hbar\Delta=0.01\epsilon$ and $\Omega=100\epsilon/\hbar$.
Figure \ref{fig:power} shows the absolute value of the heat current to the left bath for a positive temperature gradient $\Delta T=T_R-T_L=0.1$K (full line) and for a negative gradient $-0.1$K (dashed line) in function of the coupling strength $\eta_R$, with $\eta_L=\hbar$ constant.  The curves show rectification of the heat current, as was already observed by \cite{Segal2005,Segal2006}: the  current changes direction when the temperatures of the bath are exchanged, but the magnitudes are not equal. 

\begin{figure}
\includegraphics[scale=1]{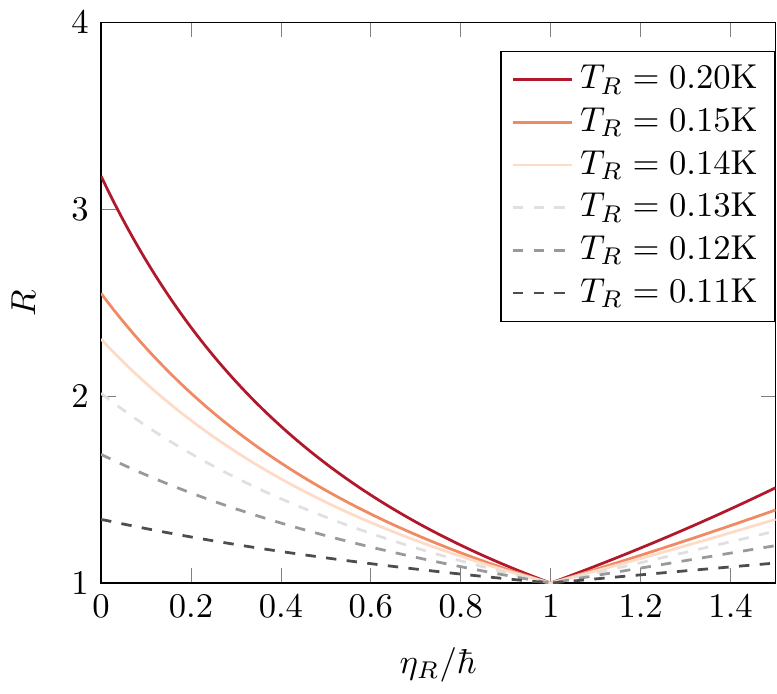}
\caption{Rectification index R, as defined in \eqref{eq:def-rect}, for different values of $T_R$ and $T_L=0.1\, \text{K}$. The other parameters are given in the caption of figure \ref{fig:power}}
\label{fig:rectification}
\end{figure}

Let $P_L$ be the power to the left bath and $P^R_L$ be the power to the left bath as the temperatures of the baths are exchanged. To quantify the amount of rectification, we define the rectification index as \cite{SeGu2019}
\begin{equation}\label{eq:def-rect}
R=\frac{\text{max}(|P_{L}|,|P^R_{L}|)}{\text{min}(|P_{L}|,|P^R_{L}|)}.
\end{equation}
The rectification index is shown in figure \ref{fig:rectification} for different range of temperatures of the right bath in function of the coupling parameter $\eta_R$. Larger temperature gradients lead to higher rectification.

\begin{figure}
\includegraphics[scale=1]{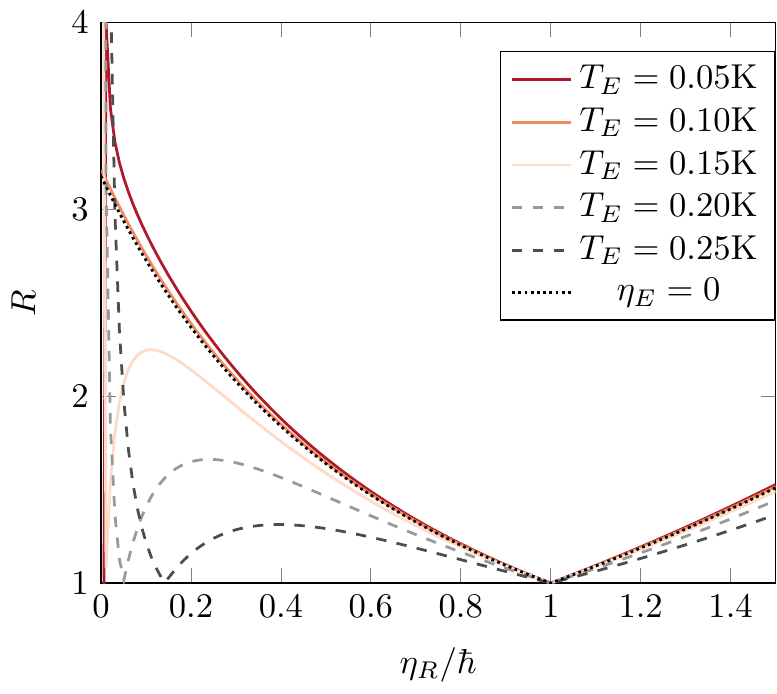}
\caption{Influence of a third bath on the rectification index $R$, as defined in  \eqref{eq:def-rect}. The coupling to the third bath is $\eta_E=0.1\hbar$, $T_L=0.1\,\text{K}$ and $T_R=0.2\,\text{K}$. The other parameters are given in the caption of figure \ref{fig:power}}
\label{fig:thirdbath2}
\end{figure}

The influence of a third bath, with temperature $T_E$, weakly coupled to the qubit on the rectification index $R$ is shown in figure \ref{fig:thirdbath2}. The left bath has constant coupling $\eta_L=\hbar$, the third bath has coupling $\eta_E=0.1\hbar$ and the coupling of the right bath ranges from $0\hbar$ to $1.5\hbar$. The presence to the third bath leads to $P_R\neq -P_L$, which causes changes in the behaviour of the rectification index $R$. The black (full) line in figure \ref{fig:thirdbath2}  displays the rectification index without the third bath, the other curves show the rectification under the influence of the third bath. There are  two clear qualitative deviations from the two-bath situation.  First, the rectification no longer reaches a minimum at $\eta_R=1$, the minima are shifted to other values of $\eta_R$ and even additional minima appear. Secondly, divergences occur when the presence of the third bath leads to $P_L=0$ and $P_L^R\neq 0$, or the other way around. For example, at $\eta_R=0$ and $T_E=0.1\, \text{K}$ the power $P_L=0$, since $T_E=T_L$ and there is no interaction with the right bath. When the temperatures are reversed, $T_E\neq T^R_L=T_R$ leading to $P^R_L\neq0$. Theoretical studies of electronic systems have show similar effect on the rectification due the influence of a third bath \cite{Snchez2017,Goury2019}, earlier numerical studies for the three bath model in the spin-boson case are \cite{Segal2008}.

\section{Discussion}
In this paper we have studied the heat current
through a qubit between two thermal baths. Earlier studies performed calculations were done using the polaron transform \cite{Segal2005,Nicolin2011,NiSe2011,Friedman2018,Wang2015,Wang2017,Liu2018}, or when explicitly performing the  non-interacting blip approximation
(NIBA) were focussed on the first moment \cite{Aurell2019}. Here we rederived the explicit expression for the generating function of the heat current
by directly performing the NIBA. The Laplace transform
of the cumulant generating function of the heat current
is a large deviation function (or rate function) that allows one to quantify rare events. In equilibrium, it can
be shown that rate functions are simply related to the
traditional thermodynamic potentials such as entropy or
free energy \cite{Touchette}. Far from equilibrium, large deviation
functions can be defined for a large class of dynamical
processes and are good candidates for playing the role of
generalized potentials  \cite{Derrida2007}. 

In classical physics, a few exact
solutions for the large deviations in some integrable interacting particles models have been found and a non-linear
hydrodynamic theory, known as macrocoscopic fluctuation theory,  has
been developed \cite{Derrida2007,Gianni2014}. In the quantum case,
the role of  large deviation
functions  is played by the full counting statistics (FCS)
\cite{BelzigNazarovPRL,BagretsNazarov,Flindt2005,Pistolesi,MaisiVille,KomnikGogolin} for which
a path integral fomulation akin to macrocoscopic fluctuation theory has been formulated \cite{Pilgram}.
The FCS exhibits universal features and phase transitions \cite{FlindtPNAS} and obeys the Fluctuation Theorem
\cite{Utsumi,SMuk1,SMuk2,SMukRV}. However,  in the quantum realm, 
exact results  for interacting systems are  very rare, amongst the  most  noticeable  is 
a series of   remarkable calculations  performed for the XXZ open spin chain interacting
with boundary reservoirs within the Lindblad framework
\cite{ProsenPRL,ProsenReview}.

In the present work, our aim was to study the heat
transport in the spin-boson model, starting from the microscopic model that embodies the qubit and the reservoirs. We hence do not rely on a Markovian assumption,
but eventually that the tunelling element is small, as is 
inherent to the NIBA.

Our analysis begins with the exact expression of the generating function in terms of a Feynman-Vernon type path
integral from which we derived a full analytical formula for
the generating function of the heat current. We recover the earlier results derived using the polaron transform \cite{Nicolin2011,NiSe2011}.

As a numerical example we studied the first moment of the generating function, the heat current. We saw that
this shows rectification when the coupling strength of
the qubit to both baths is not equal, as was already found by \cite{Segal2005,Segal2006}. When the temperature gradient is flipped, the current changes direction,
but it does not have the same magnitude in both directions
and therefore breaks the Fourier Law of heat conduction.

A very important property satisfied by the generating function   is the Gallavotti-Cohen fluctuation theorem 
that embodies at the macroscopic scale the time-reversal
invariance of the microscopic dynamics. The fluctuation
theorem implies in particular the fluctuation-dissipation
relation and the Onsager reciprocity rules when different
currents are present \cite{EvansCM,GCohen1,Gallavotti2,Gaspard2009,GaspardOnsag1}.

The fact that
the formal definition of generating function does obey the
Gallavotti-Cohen symmetry is rather straightforward to
obtain. This relation remains true after the NIBA \cite{Nicolin2011,NiSe2011}. This means that
NIBA respects the fundamental symmetries of the
underlying model, or equivalently, that the spin-boson
problem with NIBA is by itself a thermodynamically consistent
model. One consequence is that the fluctuation-dissipation relation is retrieved under the NIBA. Indeed,
we explicitly calculated the first and second moment
of the heat. When the temperature difference between the
baths is small, we found the heat conductance $\kappa$
as the first moment of heat per unit time divided by temperature difference.
The variance of the heat at equilibrium,
when both temperatures are the same, is then per unit time
proportional to $\kappa$.
We emphasize that the Gallavotti-Cohen
relation is valid far from equilibrium and it implies relations between response coefficients at arbitrary orders
 \cite{GaspardNL}. 

\section*{Acknowledgements}
\label{ack}
We gratefully acknowledge discussions with Jukka Pekola, Dmitry Golubev and Rafael Sanchez. We also thank Hugo Touchette for many useful suggestions.
This work was initiated at the Nordita program
"New Directions in Quantum Information'' (Stockholm, April 2019).
We thank Nordita, Quantum Technology Finland (Espoo, Finland), 
and International Centre for Theory of Quantum Technologies
(Gda\'nsk, Poland)
for their financial support for this event. 
The work of EA was
partially supported by Foundation for Polish Science
through TEAM-NET project (contract no. POIR.04.04.00-00-17C1/18-00).
The work of B.D. is supported by DOMAST.

\onecolumngrid
\appendix
\section{Derivation of the equation \eqref{genNobath}}\label{sec:appGenn}
The expression for the generating function \eqref{genbath} can be rewritten by defining
\begin{equation}
\bar{H}=e^{i(\alpha_RH_R+\alpha_L H_L)/\hbar}H e^{-i(\alpha_RH_R+\alpha_L H_L)/\hbar}=H_S+H_L+H_R+\bar{H}_{SL}+\bar{H}_{SR}
\end{equation}
with
\begin{equation}
\bar{H}_{SL/R}=-\sigma_z\sum_{b\in L/R}L_b\sqrt{\frac{\hbar}{2m\omega_{b, L/R}}}(b_b e^{-i\omega_b \alpha_{L/R}}+b^\dagger_b e^{i\omega_b \alpha_{L/R}}).
\end{equation}
Let $\bar{U}_{t,\alpha_R,\alpha_L}$ be the corresponding evolution operator, the generating function \eqref{genbath} is
\begin{align}
  G_{i,f}(\vec{\alpha},t)=&\tr\langle f|\bar{U}_{t,\alpha_R,\alpha_L}
  \Big(\rho_{\beta_L}\otimes\rho_{\beta_R}\otimes|i\rangle \langle i| \Big)
   \, U^\dagger_t|f\rangle,
\end{align}
With this expression the influence functional can be derived in the usual way \cite{WeissBook}, leading to \eqref{eq:influence}.
\section{Surviving terms of the NIBA}\label{app:NIBA}

It is convenient to define the Sojourn-index
\begin{equation}
\chi_t=X_t+Y_t
\end{equation}
such that during a Sojourn $X_t=Y_t=\frac{1}{2}\chi_t$
and the blip-index
\begin{equation}
\xi_t=X_t-Y_t.
\end{equation}
such that during a blip
$X_t=-Y_t=\frac{1}{2}\xi_t$.
Since we will be performing two time integrals, we will be needing the second primitive functions $K^{R/L}_i$, $K^{R/L}_r$ of $k^{R/L}_i(t-s)$ and $k^{R/L}_r(t-s)$. The second primitive is defined as
\begin{equation}
K^{R/L}_{i/r}=\int dt ds\,k^{R/L}_{i/r}(t-s)
\end{equation} Note that the primitive functions have an extra minus sign, due to the fact that we are integrating over $-s$
\begin{subequations}\label{eq:prim}
\begin{equation}
K^{R/L}_i(t)=\sum_b\frac{(C_{b, R/L})^2}{2m_b\omega^3_{b, R/L}}\sin(\omega_{b, R/L} t)
\end{equation}
\begin{equation}
K^{R/L}_r(t)=\sum_b\frac{(C_{b, R/L})^2}{2m_b\omega^3_{b, R/L}}\coth\left(\frac{\hbar\omega_{b, R/L}\beta_{R/L}}{2}\right)\cos(\omega_{b, R/L} t)
\end{equation}
\end{subequations}
\subsection{Blip-Blip}
We consider a blip interval that runs from a time $t^*$ to $t^*+\Delta t_b$. 
\paragraph{Imaginary part of the action}
Notice that in the same blip interval $X_t=X_s=-Y_t=-Y_s$, hence $X_tX_s=Y_tY_s=1$ and $X_tY_s=Y_tX_s=-1$. This means that the term proportional to $X_tX_s-Y_tY_s$ in the imaginary part of the action \eqref{eq:Si} will not contribute. The remaining terms which we denote by $R(\vec\alpha,t)=R^{R}(\alpha_R,t)+R^{L}(\alpha_L,t)$, give
\begin{align}
R^j(\alpha_j,t)-\frac{1}{2}K^j_i(\alpha_j)&=-\frac{1}{4}\int_{t^*}^{t^*+\Delta t_B}\int_{t^*}^{t^*+\Delta t_B}\diff t \diff s\,k^j_i(t-s+\alpha_j)\nonumber\\&=\frac{1}{4}(
K^j_i(\Delta t_B+\alpha_j)+K^j_i(-\Delta t_B+\alpha_j)-2K^j_i(\alpha_j))\nonumber\\
&=\frac{1}{2}\sum_b \frac{C^2_{b, j}}{2m_{b, j}\omega_{b, j}^3}\sin(\omega_b\alpha_j)\cos(\omega_{b, j}\Delta t_B)-\frac{1}{2}K^j_i(\alpha_j),
\end{align}
where $j=R$ or $L$.
We isolated the $\frac{1}{2}K^j_i(\alpha_j)$ term to anticipate a cancellation with Sojourn-Sojourn terms.
\paragraph{Real part of the action}
For the real part, all terms contribute. The result is $C(\vec\alpha,\Delta t_B)=C^R(\alpha_R,\Delta t_B)+C^L(\alpha_L,\Delta t_B)$, with
\begin{align}\label{eq:bbre}
& C^j(\alpha,\Delta t_B)\equiv\frac{1}{4}\int_{t^*}^{t^*+\Delta t_B}\int_{t^*}^{t}\diff t\diff s  (2k^j_r(t-s)+k_r(t-s+\alpha)+k_r(t-s-\alpha))\nonumber\\&=\frac{1}{4}(-2K^j_r(\Delta t_B)+2K^j_r(0)-K^j_r(\Delta t_B+\alpha)-K^j_r(\Delta t_B-\alpha)+2K^j(\alpha))\nonumber\\
&=\frac{1}{2}\sum_b\frac{C^2_{b, j}}{2m_{b, j}\omega_{b, j}^3}\coth\left(\frac{\omega_{b, j}\hbar\beta}{2}\right)[\cos(\omega_{b, j}\alpha)+1][1-\cos(\omega_{b, j}\Delta_B)]
\end{align}
\subsection{Blip-Sojourn}
We consider a blip interval running from $t^*-\Delta t_b$ to $t^*$ and the ensuing Sojourn interval from  $t^*$ to $t^*+\Delta t_s$.
\paragraph{Imaginary part}
The contribution from the imaginary part of the action is $\chi\xi X_-(\vec\alpha,\Delta t_B)$,
$X_-(\vec\alpha,\Delta t_B)=X^R_-(\alpha_R,\Delta t_B)+X^L_-(\alpha_L,\Delta t_B)$ with
\begin{align}
X^j_{-}(\alpha,\Delta t_B)&=\frac{1}{4}\int_{t^*}^{t^*+\Delta t_S}\diff t\int_{t^*-\Delta t_B}^{t^*}\diff s\, (2k^j_i(t-s)-k^j_i(t-s+\alpha)-k^j_i(t-s-\alpha))\nonumber\\
&=\frac{1}{4}\bigg(2K^j_i(\Delta t_S)-2K^j_i(\Delta t_S+\Delta t_B)+2K^j_i(\Delta t_B)-2K^j_i(0)\nonumber\\&\quad-K^j_i(\Delta t_S+\alpha)+K^j_i(\Delta t_S+\Delta t_B+\alpha)-K^j_i(\Delta t_B+\alpha)+K^j_i(\alpha)\nonumber\\&\quad-K^j_i(\Delta t_S-\alpha)+K^j_i(\Delta t_S+\Delta t_B-\alpha)-K^j_i(\Delta t_B-\alpha)+K^j_i(-\alpha)\bigg)
\end{align}
Following the NIBA, we have $K_i(\Delta t_S)=K_i(\Delta t_S+\Delta t_B)=K_i(\Delta t_S+\Delta t_B+\alpha)$, which leads to a significant simplification in the above equation, we find
\begin{align}
&X^j_{-}(\alpha,\Delta t_B)=\frac{1}{4}(2K^j_i(\Delta t_B)-K^j_i(\Delta t_B-\alpha)-K_i(\Delta t_B+\alpha))\\ 
&=\frac{1}{2}\sum\frac{C_{b, j}^2}{2m_{b, j}\omega_b^3}\sin(\omega_{b, j}\Delta t_B)[1-\cos(\omega_{b, j}\alpha)]
\end{align}
\paragraph{Real part}
The real part gives
$\chi\xi F_-(\vec\alpha,\Delta t_B)$,
$F_-(\vec\alpha,\Delta t_B)=F^R_-(\alpha_R,\Delta t_B)+F^L_-(\alpha_L,\Delta t_B)$
\begin{align}
F^j_{-}(\alpha,\Delta t_B)&=\frac{1}{4}\chi\xi\int_{t^*}^{t^*+\Delta t_S}\diff t\int_{t^*-\Delta t_B}^{t^*}\diff s\,(k^j_r(t-s+\alpha)-k^j_r(t-s-\alpha))\nonumber\\&=\frac{1}{4}\chi\xi\bigg( K^j_r(\Delta t_S+\alpha)+K^j_r(\Delta t_B+\alpha)-K^j_r(\Delta t_B+\Delta t_S+\alpha)-K^j_r(\alpha)\nonumber\\&\quad-K^j_r(\Delta t_S-\alpha)-K^j_r(\Delta t_B-\alpha)+K^j_r(\Delta t_B+\Delta t_S-\alpha)+K^j_r(-\alpha)\bigg)
\end{align}
Under the same argument as for the imaginary part, we get
\begin{align}
F^j_{-}(\alpha,\Delta t_B)&=\frac{1}{4}\bigg(K^j_r(\Delta t_B+\alpha)-K^j_r(\Delta t_B-\alpha)\bigg)\\
=&-\frac{1}{2}\sum_b\frac{(C^j_b)^2}{2m_b\omega_b^3}\coth\left(\frac{\omega_b\hbar\beta}{2}\right)\sin(\omega_b\Delta t_B)\sin(\omega_b\alpha)
\end{align}
\subsection{Sojourn-Blip}
The blip interval runs from $t^*-\Delta t_s$ to $t^*$ and the blip interval  $t^*$ to $t^*+\Delta t_b$.
\paragraph{Imaginary part}
This calculation is similar to the Blip-Sojourn term, but with less cancellations. 
\begin{align}
X^j_+(\alpha,\Delta t_b)&=\frac{1}{4}\int_{t^*}^{t^*+\Delta t_B}\diff t\int_{t^*-\Delta t_S}^{t^*}\diff s\, (2k^j_i(t-s)+k^j_i(t-s+\alpha)+k^j_i(t-s-\alpha))\nonumber\\
&=\frac{1}{4}\bigg(2K^j_i(\Delta t_S)-2K^j_i(\Delta t_S+\Delta t_B)+2K^j_i(\Delta t_B)-2K^j_i(0)\nonumber\\&\quad+K^j_i(\Delta t_S+\alpha)-K^j_i(\Delta t_S+\Delta t_B+\alpha)+K^j_i(\Delta t_B+\alpha)-K^j_i(\alpha)\nonumber\\&+K^j_i(\Delta t_S-\alpha)-K^j_i(\Delta t_S+\Delta t_B-\alpha)+K^j_i(\Delta t_B-\alpha)-K^j_i(-\alpha)\bigg)
\end{align}
Again, under NIBA, we have $K^j_i(\Delta t_S)=K^j_i(\Delta t_S)=K^j_i(\Delta t_S+\Delta t_B)K^j_i(\Delta t_S+\Delta t_B+\alpha)$, which gives
\begin{align}
X^+(\alpha,\Delta t_b)&=\frac{1}{4}(2K^j_i(\Delta t_B)+K^j_i(\Delta t_B-\alpha)+K^j_i(\Delta t_B+\alpha))\nonumber\\&=\frac{1}{2}\sum\frac{(C^j_b)^2}{2m_b\omega_b^3}\sin(\omega_b\Delta t_B)[1+\cos(\omega_b\alpha)].
\end{align}
\paragraph{Real part}
\begin{align}
F^j_+(\alpha,\Delta t_b)&=\frac{1}{4}\chi\xi\int_{t^*}^{t^*+\Delta t_S}\diff t\int_{t^*-\Delta t_B}^{t^*}\diff s\,(-k^j_r(t-s+\alpha)+k^j_r(t-s-\alpha))\nonumber\\&=\chi\xi\frac{1}{4}\bigg( -K^j_r(\Delta t_S+\alpha)-K^j_r(\Delta t_B+\alpha)+K^j_r(\Delta t_B+\Delta t_S+\alpha)+K^j_r(\alpha)\nonumber\\&+K^j_r(\Delta t_S-\alpha)+K^j_r(\Delta t_B-\alpha)-K^j_r(\Delta t_B+\Delta t_S-\alpha)-K^j_r(-\alpha)\bigg)
\end{align}
Under the same argument as for the imaginary part, we get
\begin{align}
F^j_+(\alpha,\Delta t_b)&=\frac{1}{4}\bigg(-K^j_r(\Delta t_B+\alpha)+K^j_r(\Delta t_B-\alpha)\bigg)\nonumber\\&=\frac{1}{2}\sum\frac{(C_b^j)^2}{2m_b\omega_b^3}\coth\left(\frac{\omega_b\hbar\beta}{2}\right)\sin(\omega_b\Delta t_B)\sin(\omega_b\alpha).
\end{align}
Note that $F_+=-F_-$.
\subsection{Sojourn-Sojourn}
The  first Sojourn interval runs from $t^*$ to $t^*+\Delta t_{S_1}$ and the blip interval  $t^*+\Delta t_{S_1}$ to $t^*+\Delta t_{S_1}+\Delta t_{S_2}$.
\paragraph{Imaginary part}
We find
\begin{align}
B^j(\alpha)\equiv&\frac{1}{4}\int_{t^*}^{t^*+\Delta t_S}\int_{t^*}^{t^*+\Delta t_S}\diff t \diff s\,k^j_i(t-s+\alpha)\nonumber\\&=\frac{1}{4}(2K^j_i(\alpha)-K^j_i(\Delta t_S+\alpha)-K^j_i(-\Delta t_S+\alpha))\nonumber\\&=\frac{1}{2}K^j_i(\alpha)\nonumber\\&=\frac{1}{2}\sum_b\frac{(C^j_b)^2}{2m_b\omega_b^3}\sin(\alpha)
\end{align}
\paragraph{Real part}
\begin{align}
D^j(\alpha)&=\frac{1}{4}\int_{t^*}^{t^*+\Delta t_S}\diff t\int_{t^*}^{t}\diff s \,(2k^j_r(t-s)-k^j_r(t-s+\alpha)-k^j_r(t-s-\alpha))\nonumber\\&=\frac{1}{4}\bigg(2K^j_r(0)-2K^j_r(\Delta t_S)+K^j_r(\Delta t_S +\hbar \alpha)-K^j_r(\alpha)+K^j_r(\Delta t_S -\hbar \alpha)-K^j_r(-\alpha)\bigg)\nonumber\\&=\frac{1}{2}(K^j_r(0)-K^j_r(\alpha))\\&=\frac{1}{2}\sum\frac{(C_b^j)^2}{2m_b\omega_b^3}\coth\left(\frac{\omega_b\hbar\beta_j}{2}\right)[1-\cos(\alpha)]
\end{align}
There will also be cancellations between $D$ and $C$.
\subsection{Sojourn-(Blip)-Sojourn}
The  first Sojourn interval runs from $t^*$ to $t^*+\Delta t_{S_1}$ and the second Sojourb interval from $t^*+\Delta t_{S_1}+\Delta t_{B}$ to $t^*+\Delta t_{S_1}+\Delta t_{B_1}+\Delta t_{S_2}$, where $\Delta t_{B}$ is the duration of the blip.
\paragraph{Imaginary part}
\begin{align}
\Lambda^j(\alpha,\Delta t_B)&=\frac{1}{4}\int_{t^*+\Delta t_{S_1}+\Delta t_{B}}^{t^*+\Delta t_{S_1}+\Delta t_{B}+\Delta t_{S_2}}\diff t \int_{t^*}^{t^*+\Delta t_{S_1}}\diff s\,(k^j_i(t-s+\alpha)-k^j_i(t-s-\alpha))\nonumber\\&=\frac{1}{4}(-K^j_i(\Delta t_B+\alpha)+K^j_i(\Delta t_B-\alpha))\nonumber\\&=-\frac{1}{2}
\sum_b\frac{(C_b^j)^2}{2m_b\omega_b^3}\cos(\omega_b\Delta t_B)\sin(\omega_b\alpha)
\end{align}
\paragraph{Real part}
\begin{align}
\Sigma^j(\alpha,\Delta t_B)&=\frac{1}{4}\int_{t^*+\Delta t_{S_1}+\Delta t_{B}}^{t^*+\Delta t_{S_1}+\Delta t_{B}+\Delta t_{S_2}}\diff t \int_{t^*}^{t^*+\Delta t_{S_1}}\diff s\,(2k^j_r(t-s)-k^j_r(t-s+\alpha)-k^j_r(t-s-\alpha))\nonumber\\
&=\frac{1}{4}(K^j_r(\Delta t_B+\alpha)+K^j_r(\Delta t_B-\alpha)-2 K^j_r(\Delta t_B))\nonumber\\
&=\frac{1}{2}\sum\frac{(C_{b, j})^2}{2m_{b, j}\omega_{b, j}^3}\coth\left(\frac{\omega_{b, j}\hbar\beta_j}{2}\right)\cos(\omega_{b, j}\Delta t_B)[\cos(\omega_{b, j}\alpha)-1]
\end{align}

\subsection{Transfer matrix}
The generating function, using the terms calculated in the last subsections, is
\begin{align}\label{eq:actSS}
&G_{S\rightarrow S}(\alpha)=\sum_{n=0}^{+\infty}(-1)^n\left(\frac{\Delta}{2}\right)^n\int \diff t_1\hdots \diff t_{2n}\,\sum_{\substack{\chi_1,\hdots,\chi_n=\pm 1, \\\xi_1,\hdots,\xi_n=\pm 1}}\exp\bigg(-\frac{i}{\hbar}\epsilon\sum_i\xi_{i}(t_{2i}-t_{2i-1})\bigg)\nonumber\\
&\times\exp\bigg(\frac{i}{\hbar}\sum_{j=R,L}\sum_i\chi_{i}\xi_{i+1}X^j_+(\alpha_j,\Delta_{2i+2})+\chi_i\xi_i X^j_-(\alpha_j,\Delta_{2i})+\chi_i\chi_{i+1}\Lambda^j(\alpha,\Delta_{2i+2})+R^j(\alpha,\Delta_{2i})\bigg)\nonumber\\
&\times\exp\bigg(-\frac{1}{\hbar}\sum_{j=R,L}\sum_i\chi_{j}\xi_{i+1}F^j_+(\alpha_j,\Delta_{2i+2})+\chi_i\xi_i F^j_-(\alpha_j,\Delta_{2i})+\chi_i\chi_{i+1}\Sigma^j(\alpha_j,\Delta_{2i+2})+C'(\alpha_j,\Delta_{2i})\bigg)
\end{align}
To express the resulting generating function in terms of a transfer matrix, it is convenient to first define for $j=R$ or $L$
\begin{subequations}
\begin{equation}
Z^+_{j}(t)=X^j_+(\alpha,t)+X^j_-(\alpha,t)=\frac{2\eta_j}{\pi}\int_0^\Omega\diff \omega\, \frac{1}{\omega}\sin(\omega t)
\end{equation}
\begin{equation}
Z_j^-(\alpha,t)=X^j_+(\alpha,t)-X^j_-(\alpha,t)=\frac{2\eta_j}{\pi}\int_0^\Omega\diff \omega\, \frac{1}{ \omega}\sin(\omega t)\cos(\omega\alpha)
\end{equation}
\end{subequations}
and
\begin{subequations}
\begin{align}
&\Gamma^+_j(t)=C^j(\alpha,t)+D^j(\alpha,t)+\Sigma(\alpha,t)=\frac{2\eta_j}{\pi}\int_0^\Omega\diff \omega\, \frac{1}{\omega}\coth\left(\frac{\omega\hbar\beta_j}{2}\right)(1-\cos(\omega t))
\end{align}
\begin{align}
&\Gamma^-_j(\alpha,t)=C^j(\alpha,t)+D^j(\alpha,t)-\Sigma(\alpha,t)=\frac{2\eta_j}{\pi}\int_0^\Omega\diff \omega\, \frac{1}{ \omega}\coth\left(\frac{\omega\hbar\beta_j}{2}\right)(1-\cos(\omega t)\cos(\omega\alpha))
\end{align}
\end{subequations}
which allows us to write the generating function as \eqref{generatinTransf}.

\section{Heat current}
\label{sec:thermal-power}
In this section we are interested in studying the heat current between the two baths. The heat current is defined as 
\begin{equation}
\label{eq:thermal-power}
\Pi(\beta_C,\beta_R) = \lim_{t\to\infty}\frac{\left<\Delta E_c\right>}{t}
\end{equation}
where $\beta_L$ and $\beta_R$ are the inverse temperatures of respectively the
left bath and the right bath, and $E_c$ is the energy of the left bath.

To our knowledge the results in this section were first obtained in~\cite{SeNi2005}
although only stated for the case of zero level splitting.
Here we we-derive them using the same notation as in the main body of the paper.
We will show that $\Pi(\beta,\beta)=0$, which one would physically expect. It means that in the steady state there is no heat transfer between two baths  with the same temperature.
Furthermore, we calculate the thermal conductance $\kappa$ which is defined by the expansion for small temperature differences $\Delta\beta$ in both baths
\begin{equation}
\Pi(\beta,\beta+\Delta\beta) = \kappa\Delta\beta +O(\Delta\beta^2).
\end{equation}

Our starting point is a result by the authors of \cite{Aurell2019} for the form of the heat current
\begin{equation}
\label{eq:thermal-power-AM}
\Pi =(\frac{\Delta}{2})^2 \left( \frac{p_-}{p_++p_-}\pi_{\uparrow} + \frac{p_+}{p_++p_-}\pi_{\downarrow} \right),
\end{equation}
where $\frac{p_-}{p_++p_-}$ is the steady state population of the lower qubit state and $(\frac{\Delta}{2})^2\pi_{\uparrow}$ the heat current related to this state.

Let us introduce the characteristic functions
\begin{eqnarray}
C_L(t) &=& e^{-\frac{1}{\hbar}\Gamma_L^+(t)+ \frac{i}{\hbar}Z^+_L(t)} \\ 
C_R(t) &=& e^{-\frac{1}{\hbar}\Gamma_R^+(t)+ \frac{i}{\hbar}Z^+_R(t)},
\end{eqnarray}
which allows us to conveniently write the coefficients of \eqref{eq:thermal-power-AM}
\begin{subequations}
\begin{align}
p_+&= \int_{-\infty}^\infty dt\,C_L(t)C_R(t)e^{i\epsilon t}\\
p_-&= \int_{-\infty}^\infty dt\,C_L(t)C_R(t)e^{-i\epsilon t},
\end{align}
\end{subequations}
\begin{subequations}
\begin{align}
\pi_{\uparrow}&=-{i\hbar}\int_{-\infty}^\infty dt\,\frac{dC_L(t)}{dt}C_R(t)e^{i\epsilon t}\\
\pi_{\downarrow}&=-i{\hbar}\int_{-\infty}^\infty dt\,\frac{dC_L^+(t)}{dt}C_R^+(t)e^{-i\epsilon t},
\end{align}
\end{subequations}
and
\begin{subequations}
\begin{align}
\Sigma^+&=-\hbar^2\int_{-\infty}^\infty dt\,\frac{d^2C_L(t)}{dt^2}C_R(t)e^{i\epsilon t}\\
\Sigma^-&=-\hbar^2\int_{-\infty}^\infty dt\,\frac{d^2C_L(t)}{dt^2}C_R(t)e^{-i\epsilon t}
\end{align}
\end{subequations}
\subsection{Two baths with the same temperatures}
When both baths have the same temperatures, we expect the steady state heat transfer to be zero
\begin{equation}
\label{eq:thermal-power-zero}
\Pi(\beta,\beta) = 0
\end{equation}

Via an analytic continuation argument outlined in Appendix \ref{sec:analytical}, we find that
 \begin{align}\label{eq:analcont1}
 p_+(\beta_L,\beta_R) = &\frac{1}{2}\int d t C_L(t+i\Delta\beta\hbar) C_R(t) e^{-\frac{i}{\hbar}\epsilon t} e^{-\epsilon\beta_R} 
\end{align}
and
 \begin{align}\label{eq:analcont2}
\pi_\downarrow(\beta_C,\beta_R) &={i\hbar} \int d t \frac{d C_L(t+i\Delta\beta\hbar)}{dt} C_R(t) e^{\frac{i}{\hbar}\epsilon t} e^{\epsilon\beta_R} 
\end{align}
with $\Delta\beta=\beta_C-\beta_R$.
When both temperatures are equal, these relations transform to
\begin{eqnarray}
\label{eq:D-def-4}
p_+(\beta,\beta) 
               &=&  e^{-\epsilon\beta} p_-(\beta,\beta)
\end{eqnarray}
and
\begin{eqnarray}
\label{eq:pi}
\pi_{\uparrow}(\beta,\beta)  &=& - e^{-\epsilon\beta}\pi_{\downarrow}(\beta,\beta).
\end{eqnarray}
Equations \eqref{eq:D-def-4} and \eqref{eq:pi} directly give us 
\begin{eqnarray}
\label{eq:thermal-power-AM-same}
\Pi(\beta,\beta) &=& \left(\frac{\Delta}{2}\right)^2 \frac{1}{p_++p_-}\left(p_-\pi_{\uparrow} + p_+\pi_{\downarrow}\right) \nonumber \\
 &=&\left(\frac{\Delta}{2}\right)^2 \frac{e^{-\epsilon\beta}}{p_++p_-}  \left(-p_-\pi_{\downarrow} +p_-\pi_{\downarrow}\right) = 0
\end{eqnarray}
\subsection{Thermal Conductance}
\label{sec:almost-same-temperatures}
The obtain an explicit formula for the thermal conductance $\kappa$ one should expand \eqref{eq:thermal-power-AM} in the difference between the temperature of both baths $\Delta\beta=\beta_L-\beta_R$.
Differentiating the denominator ($A+D$) gives no contribution
as it multiplies a parenthesis $D\pi_{\uparrow} + A\pi_{\downarrow}$, which vanishes to zeroth order.
We can therefore write
\begin{eqnarray}
\label{eq:thermal-power-AM2}
\kappa &=&\left(\frac{\Delta}{2}\right)^2 \frac{1}{p_++p_-}\Big(
\partial_{\beta_L}(p_-)\pi_{\uparrow} + p_- \partial_{\beta_L}(\pi_{\uparrow}) \nonumber \\
&& \quad + \partial_{\beta_L}(p_+)\pi_{\downarrow} + p_+ \partial_{\beta_L}(\pi_{\downarrow})\Big)
\end{eqnarray}
All terms on the right hand side are evaluated at $\beta_L=\beta_R=\beta$.

The calculation of $\kappa$ is presented in Appendix \ref{sec:appTherm}.
The idea of the calculation is to write out $\partial_{\beta}(p_-)\pi_{\uparrow}$
and $p_- \partial_{\beta}(\pi_{\uparrow})$, and to keep track how the
terms generated in the partial derivatives  $\partial_{\beta}(p_-)$ and
and $\partial_{\beta}(\pi_{\uparrow})$ change as the 
integral variable $t$ is shifted to
$t+i\hbar\beta$. 
The result is 
\begin{align}\label{eq:kappafinal}
&\kappa=\left(\frac{\Delta}{2}\right)^2\frac{1}{p_++p_-}({p_+\Sigma^-+4\pi_\downarrow\pi_\uparrow+p_-\Sigma^+}),
\end{align}
where $\tilde{C}$ and $\tilde{D}$ are the Laplace transforms of the matrix elements defined in equation \eqref{eq:def-matrixelements} and the accent denotes the derivative to the first variable.

\section{Analytic continuation}
\label{sec:analytical}
Suppose that all functions are analytical in the strip
$0\leq \Im t \leq \hbar\beta_R$  (note that  in Appendix E of  \cite{Leggett1987}
 the authors assume an analytic continuation to negative
  imaginary values of $t$; however  they consider  the function $G(t)$
  related to the function $C(t)$ by  $G(t)=e^{-C(t)}$).
Then any of the integrals, say $D$, can be written as
\begin{eqnarray}
\label{eq:D-def-2}
p_+ &=& \int d t C_L(t+i\beta_R\hbar) C_R(t+i\beta_R\hbar) e^{\frac{i}{\hbar}\epsilon (t+i\beta_R\hbar)} 
\end{eqnarray}

The exponents in $C_L$ and $C_R$ are sums over bath oscillators. Each oscillator $b$ contributes
\begin{eqnarray}
\hbox{Term} &=& \frac{1}{2m_b\omega_b}\left(-\coth\frac{\omega_b\hbar\beta}{2}(1- \cos \omega_b t) +i\cdot \sin \omega_b t\right)
\end{eqnarray}
where $\beta$ is $\beta_R$ or $\beta_L$. 
Evaluating first oscillators in the right bath gives
\begin{eqnarray}
\label{eq:cos-t}
\cos\omega_b \left(t +i\hbar\beta\right) &=& \cos\omega_b t \cosh\hbar\omega_b\beta  
-i\sin\omega_b t \sinh\hbar\omega_b\beta \\  
\label{eq:sin-t}
\sin\omega_b \left(t +i\hbar\beta\right) &=& \sin\omega_b t \cosh\hbar\omega_b\beta  
+i\cos\omega_b t \sinh\hbar\omega_b\beta   
\end{eqnarray}
which with $\coth\frac{\omega_b\hbar\beta_R}{2}$ from above can be combined into
\begin{eqnarray}
\label{eq:cos-t-1}
\cos\omega_b t\left(\coth\frac{\omega_b\hbar\beta_R}{2}  \cosh\hbar\omega_b\beta_R - \sinh\hbar\omega_b\beta_R\right)
&=& \cos\omega_b t \coth\frac{\omega_b\hbar\beta_R}{2} \\
\label{eq:sin-t-1}
i\sin\omega_b t \left(-\coth\frac{\omega_b\hbar\beta_R}{2}  \sinh\hbar\omega_b\beta_R + \cosh\hbar\omega_b\beta_R\right)
&=& - i\sin\omega_b t
\end{eqnarray}
Hence
\begin{eqnarray}
C_R(t+i\beta_R\hbar) &=& \overline{C_R(t)}= C_R(-t)
\end{eqnarray}

For the oscillators in the left bath we consider ($\Delta\beta=\beta_L-\beta_R$)
\begin{eqnarray} 
C_L(t+i\beta_R\hbar) &=& C_L(t-i\hbar\Delta\beta+i\beta_L\hbar) = \overline{L_C(t-i\hbar\Delta\beta)}= C_L(-t+i\hbar\Delta\beta)
\end{eqnarray}
Inserting back into the expression for $D$ this means
\begin{eqnarray}
\label{eq:D-def-3}
p_+(\beta_L,\beta_R) &=& \int d t C_L(-t+i\Delta\beta\hbar) C_R(-t) e^{\frac{i}{\hbar}\epsilon t} e^{-\epsilon\beta_R} \nonumber \\ 
\end{eqnarray}

\section{Thermal Conductance}
\label{sec:appTherm}
\subsubsection{The partial derivative of $D$}
\label{sec:D-term}
For $p_+$ one finds
\begin{eqnarray}
\label{eq:D-def-5}
\partial_{\beta_L}p_+ &=& \int d t \partial_{\beta_L}\left(\log  C_L(t)\right)_{\beta_L=\beta}  C_L(t)C_R(t) e^{\frac{i}{\hbar}\epsilon t}  
\end{eqnarray}
where
\begin{eqnarray}
\partial_{\beta_C}\log C_L(t)&=& \sum_{b\in C} \frac{1}{2m_b\omega_b}(1-\cos\omega_b t)\frac{1}{\sinh^2\frac{\omega_b\hbar\beta_C}{2}}\frac{\omega_b\hbar}{2} \nonumber 
\end{eqnarray}
Changing $t$ to
$t+i\hbar\beta$ will change  $C_L(t)C_R(t) e^{\frac{i}{\hbar}\epsilon t}$ to
 $C_L(-t)C_R(-t) e^{-\frac{i}{\hbar}\epsilon (-t)}e^{-\epsilon\beta}$, similarly as in Appendix \ref{sec:analytical}.

The logarithmic derivative on the other hand changes in the convenient way: 
\begin{eqnarray}
\label{eq:D-def-6}
\partial_{\beta_L}\log C_L(t+i\hbar\beta;\beta_L=\beta)&=& 
\sum_{b\in L} \frac{\omega_b\hbar}{4m_b\omega_b}(1-\cos\omega_b t)\frac{1}{\sinh^2\frac{\omega_b\hbar\beta}{2}} + \sum_{b\in L} \frac{\omega_b\hbar}{4m_b\omega_b}\cos\omega_b t (-2)  \nonumber \\ 
&+& \sum_{b\in L} \frac{\omega_b\hbar}{4m_b\omega_b}\sin\omega_b t (2i) \coth\frac{\omega_b\hbar\beta}{2}
\end{eqnarray}
The two last terms can be compared to
\begin{eqnarray}
\label{eq:comparison}
\partial_{t}\log C_L(t)&=& 
\partial_{t}\left(\sum_{b\in L} \frac{1}{2m_b\omega_b}\left[-(1-\cos\omega_b t)\coth\frac{\omega_b\hbar\beta}{2}
+i\sin\omega_b t\right]\right)\nonumber \\ 
&=& \sum_{b\in L} \frac{1}{2m_b\omega_b}\left[(-\omega_b\sin\omega_b t)\omega_b\coth\frac{\omega_b\hbar\beta}{2}
+i\omega_b\cos\omega_b t\right]
\end{eqnarray}
Eq.~\eqref{eq:D-def-6} can therefore be rewritten as
\begin{eqnarray}
\label{eq:D-def-7}
\partial_{\beta_L}\log C_L(t+i\hbar\beta;\beta_L=\beta)&=& \partial_{\beta_L}\log C_L(-t;\beta_L=\beta) + i\hbar \partial_{s}\log C_L(s;\beta_L=\beta)|_{s=-t}
\end{eqnarray}
We can now change the integral variable from $t$ to $-t$ which gives
\begin{eqnarray}
\label{eq:D-def-8}
\partial_{\beta}(p_+)\pi_{\downarrow} &=& - \partial_{\beta}(p_-)\pi_{\uparrow} -{\hbar^2} \int d t \partial_t\left(C_L(t)\right) C_R(t) e^{\frac{i}{\hbar}\epsilon t} \int d t \partial_t\left(C_L(t)\right) C_R(t) e^{-\frac{i}{\hbar}\epsilon t} 
\end{eqnarray}
The sign is determined as follows: 
$\pi_{\uparrow}$ changes sign when it goes to $\pi_{\downarrow}$,
but $D$ does not. There is factor $i\hbar$ in the definition of 
$\pi_{\uparrow}$ and another one in the second term in $\partial_{\beta_L}\log C_L(t+i\hbar\beta)$.
Taken together this gives 
$-(i\hbar) (-i\hbar)=-\hbar^2$.

\subsubsection{The partial derivative of $\pi_{\uparrow}$}
\label{sec:pi-up-term}
This term can be evaluated in practically the same way as the other one.
One starts from 
\begin{eqnarray}
\partial_{\beta}\pi_{\downarrow} &=-i& \partial_{\beta}\left(\int\cdots \partial_t(C_L(t))\cdots\right) \nonumber \\
               &=-i&  \left(\int\cdots C_L(t)\partial_{\beta}(\log C_L(t)) \partial_t(\log C_L(t))\cdots\right)  \nonumber \\
               && -i \left(\int\cdots C_L(t)\partial^2_{t\beta}(\log C_L(t))\cdots\right) 
\end{eqnarray}
One now treats $\partial_{\beta}(\log C_L(t))$ in the same way as in
\eqref{eq:D-def-7}.
The first term will then give something proportional to $(\partial_t(\log C_L(t)))^2$
and the second something proportional to $\partial_{tt}(\log C_L(t))$.
Combining we have
\begin{equation}
C_L\left((\partial_t(\log C_L(t)))^2+\partial_{tt}(\log C_L(t))\right)=\partial_{tt}C_L
\end{equation}
This means that we can write
\begin{eqnarray}
\label{eq:D-def-8-bis}
p_+\partial_{\beta}(\pi_{\downarrow}) &=& - p_-\partial_{\beta}(\pi_{\uparrow})  -{\hbar^2 }\int d t \partial_{tt}\left(C_L(t)\right) C_R(t) e^{\frac{i}{\hbar}\epsilon t} \int d t C_L(t) C_R(t) e^{-\frac{i}{\hbar}\epsilon t} 
\end{eqnarray}
The sign is determined as follows: 
$\pi_{\downarrow}$ changes sign when it goes to $\pi_{\uparrow}$,
but the terms with two time derivates do not change sign.
The factors $i\hbar$ and $-i\hbar$  are the same as before.
\subsubsection{Combination}
\label{sec:combination}
Inserting \eqref{eq:D-def-8} and \eqref{eq:D-def-8-bis}, using that 
\begin{equation}
\frac{1}{2}\int d t \partial_{tt}\left(C_L(t)\right) C_R(t) e^{\frac{i}{\hbar}\epsilon t} =\tilde{C}''(0,0)
\end{equation}
\begin{equation}
\frac{1}{2}\int d t \partial_{tt}\left(C_L(t)\right) C_R(t) e^{\frac{-i}{\hbar}\epsilon t} =\tilde{B}''(0,0)
\end{equation}
and symmetrizing
one has
\begin{equation}
\kappa=-\frac{(\hbar\Delta)^2}{4(p_++p_-)}(p_+\tilde{B}''(0,0)+4\tilde{B}'(0,0)\tilde{C}'(0,0)+p_-\tilde{C}''(0,0))
\end{equation}
\twocolumngrid

\bibliography{lit} 
\end{document}